\documentclass[12pt,draftcls,onecolumn]{IEEEtran}
\usepackage{setspace}
\ifCLASSINFOpdf
\else
\fi
\usepackage{amsmath,amsfonts}
\usepackage{algorithmic}
\usepackage{algorithm}
\usepackage{array}
\usepackage[caption=false,font=normalsize,labelfont=scriptsize,textfont=scriptsize]{subfig}
\usepackage{textcomp}
\usepackage{stfloats}
\usepackage{url}
\usepackage{verbatim}
\usepackage{graphicx}
\usepackage{subfig}
\usepackage{cite}
\usepackage{color}
\usepackage{amssymb}
\usepackage{siunitx}
\usepackage{threeparttable}
\usepackage{amsmath}
\usepackage{amssymb}
\usepackage{graphicx} 
\usepackage{subfig}
\usepackage[export]{adjustbox}
\begin{document}
\title{Channel Estimation for Stacked Intelligent Metasurface-Assisted Wireless Networks}
\author{Xianghao~Yao,
        Jiancheng~An, {\emph{Member, IEEE}},
        Lu Gan,\\ Marco Di Renzo, {\emph{Fellow, IEEE}},
        and Chau Yuen, {\emph{Fellow, IEEE}}\vspace{-1.2cm}
\thanks{
This work is partially supported by Sichuan Science and Technology Program under Grant 2023YFSY0008 and 2023YFG0291, and partially supported by Yibin Science and Technology Program under Grant YBP-002. The work of M. Di Renzo is supported in part by the Horizon Europe projects COVER-101086228, UNITE-101129618, and INSTINCT-101139161, and the ANR projects NF-PERSEUS 22-PEFT-004 and PASSIONATE ANR-23-CHR4-0003-01. The research of Chau Yuen is supported by the Ministry of Education, Singapore, under its MOE Tier 2 (Award number MOE-T2EP50220-0019).\par
X. Yao and L. Gan are with the School of Information and Communication Engineering, University of Electronic Science and Technology of China, Chengdu, Sichuan, 611731, China. L. Gan is also with the Yibin Institute of UESTC, Yibin 644000, China (e-mail: xianghao\_yao@163.com; ganlu@uestc.edu.cn). J. An and C. Yuen are with the School of Electrical and Electronics Engineering, Nanyang Technological University, Singapore 639798 (e-mail: jiancheng\_an@163.com; chau.yuen@ntu.edu.sg). M. Di Renzo is with Universit\'e Paris-Saclay, CNRS, CentraleSup\'elec, Laboratoire des Signaux et Syst\`emes, 3 Rue Joliot-Curie, 91192 Gif-sur-Yvette, France (e-mail: marco.di-renzo@universite-paris-saclay.fr).
}
}
\maketitle
\markboth{DRAFT}%
{Shell \MakeLowercase{\textit{et al.}}: Bare Demo of IEEEtran.cls for IEEE Journals}
\begin{abstract}
Emerging technologies, such as holographic multiple-input multiple-output (HMIMO) and stacked intelligent metasurface (SIM), are driving the development of wireless communication systems. Specifically, the SIM is physically constructed by stacking multiple layers of metasurfaces and has an architecture similar to an artificial neural network (ANN), which can flexibly manipulate the electromagnetic waves that propagate through it at the speed of light. This architecture enables the SIM to achieve HMIMO precoding and combining in the wave domain, thus significantly reducing the hardware cost and energy consumption. In this letter, we investigate the channel estimation problem in SIM-assisted multi-user HMIMO communication systems. Since the number of antennas at the base station (BS) is much smaller than the number of meta-atoms per layer of the SIM, it is challenging to acquire the channel state information (CSI) in SIM-assisted multi-user systems. To address this issue, we collect multiple copies of the uplink pilot signals that propagate through the SIM. Furthermore, we leverage the array geometry to identify the subspace that spans arbitrary spatial correlation matrices. Based on partial CSI about the channel statistics, a pair of subspace-based channel estimators are proposed. Additionally, we compute the mean square error (MSE) of the proposed channel estimators and optimize the phase shifts of the SIM to minimize the MSE. Numerical results are illustrated to analyze the effectiveness of the proposed channel estimation schemes.                       
\end{abstract}
\vspace{-0.25cm}
\IEEEpeerreviewmaketitle
\vspace{-0.25cm}
\section{Introduction}
\vspace{-0.05cm}
\label{Introduction}
\IEEEPARstart{N}{ext} generation wireless networks promise to provide ten-fold transmission rate improvement. Therefore, there is an urgent need for advanced wireless technologies to support ultra-reliability, low latency, and high data rate \cite{arXiV_2023_An_Toward, wang2023road,tang2022roadmap}. Among the emerging technology candidates, holographic multiple-input multiple-output (HMIMO) technology has drawn increasing interest \cite{an2023tutorial}. By integrating a large number of radiating and sensing elements, the energy can be focused into more precise areas to achieve high energy and spectral efficiencies \cite{an2023tutorial,pizzo2022fourier}. However, HMIMO is still in its initial stage of research and there is still a gap between theoretical performance and experimental validation. To address this issue, metasurface technology is a promising candidate for realizing HMIMO due to its low power consumption and reduced number of radio frequency (RF) chains \cite{arXiV_2023_An_Stacked_MAG, liu2022programmable,deng2023reconfigurable,huang2020holographic}.

To further improve the flexibility of metasurfaces for generating diverse RF waveforms, the authors of \cite{Stacked1} proposed a novel stacked intelligent metasurface (SIM) technology, which realizes advanced signal processing capabilities thanks to an architecture that resembles a deep neural network in the wave domain. Leveraging this advanced architecture, the authors of \cite{an2023stacked, arXiV_2023_An_Stacked_MU} designed an SIM-assisted HMIMO system that implements transmit precoding and receiver combining as the transmitted wave propagates through the SIM, thereby decreasing the number of RF chains. Also, an SIM greatly reduces the power consumption, precoding delay, and hardware cost compared to its digital counterpart \cite{Stacked1}. However, an SIM cannot efficiently probe the channels of each passive meta-atom, as the number of RF chains is smaller than the number of meta-atoms. To fully exploit the potential gains brought by SIM-assisted HMIMO systems, the accurate estimation of the channels becomes crucial. Furthermore, to effectively reduce the multi-user interference, the number of meta-atoms is generally larger than the number of antennas at the base station (BS) \cite{Stacked1}. This leads to an increase in the dimension of the channel matrix and poses great challenges for channel estimation in SIM-assisted HMIMO systems \cite{peng2023channel}.

Against this background, in this letter, we investigate the channel estimation in SIM-assisted multi-user HMIMO communication systems. Specifically, we propose an estimation protocol by accumulating the received signals over multiple time blocks to tackle the underdetermined channel estimation problem. By transmitting orthogonal pilot sequences to the BS in each block, the interference between multiple users can be mitigated. Based on the proposed model, we then present four distinct channel estimation techniques, including the least squares (LS), the minimum mean square error (MMSE), the reduced-subspace LS (RSLS), and the RSLS relying on isotropic scattering statistics (RSLS-iso). The latter two methods exhibit enhanced estimation accuracy even with limited knowledge of the spatial correlation matrix. Furthermore, the phase shifts of the meta-atoms in the SIM are optimized for improving the performance of the considered channel estimators. Extensive simulation results are illustrated to prove the effectiveness of the proposed channel estimation techniques.\par
\begin{figure*}[!htp]    
\centering
\subfloat[An SIM-assisted multi-user HMIMO system.]  
{\label{Figure1a}\includegraphics[height=3.7cm]{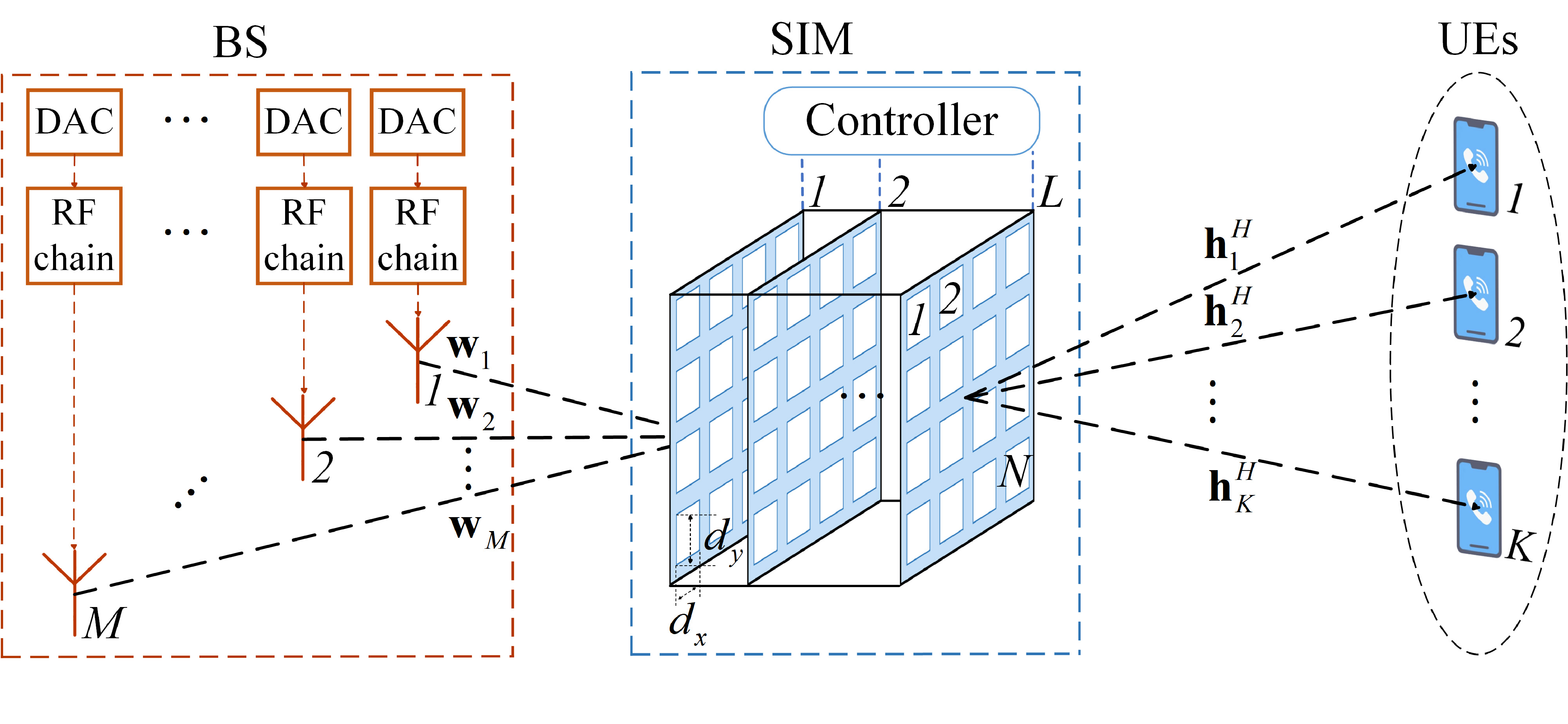}\hspace{40mm}}
\subfloat[\scriptsize Bird eye view of BS and SIM.]
{\label{Figure1b}\includegraphics[height=3.7cm]{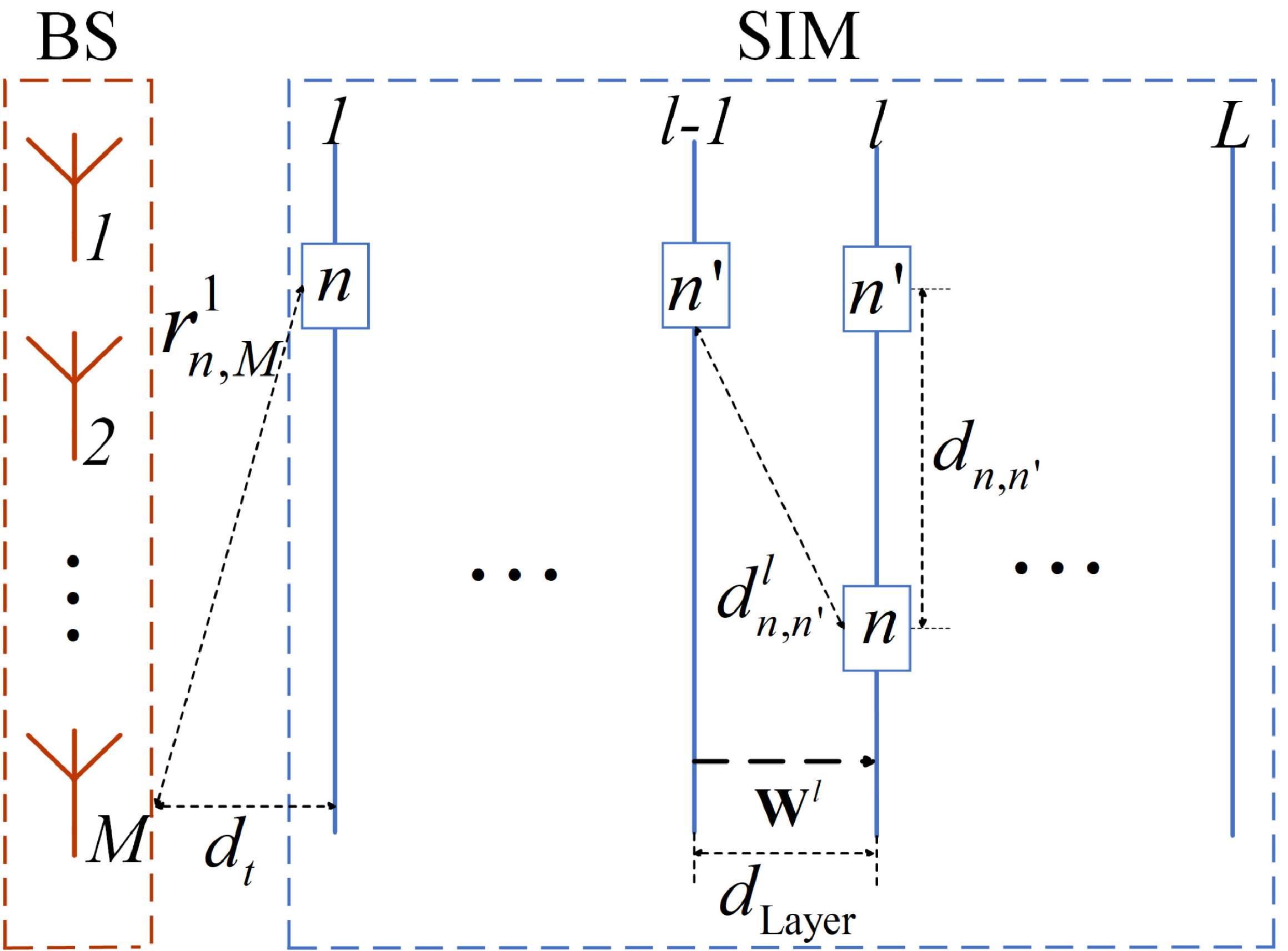}\hspace{20mm}}
\caption{SIM-assisted HMIMO transmission.}   
\label{Figure1}  
\vspace{-0.65cm}
\end{figure*}
\vspace{-0.25cm}
\section{System and Channel Models}
\vspace{-0.05cm}
\label{SYSTEM}
In this section, we introduce an SIM-assisted multi-user HMIMO communication system. As shown in Fig. \ref{Figure1a}, a BS equipped with a uniform linear array (ULA) having $M$ antennas serves $K$ single-antenna user equipments (UEs). An SIM consists of $L$ metasurface layers, each with $N$ meta-atoms. By adjusting the electromagnetic (EM) properties of all the metasurface layers with an intelligent controller, the SIM can perform wave-based processing \cite{Stacked1}.\par
Let $\mathcal{M} =\left \{ 1,2,\cdots, M  \right \} $, $\mathcal{L} = \left \{1,2,\cdots, L \right \}$, $\mathcal{N} =\left \{1,2,\cdots, N \right \}$, and $\mathcal{K} =\left \{1,2,\cdots, K \right \}$ denote the indices of the antennas, metasurface layers, meta-atoms on each layer, and UEs, respectively. The distance between the $m$-th antenna and the $n$-th meta-atom of the first metasurface layer is denoted by $r_{n,m}^{1}$. The distance between adjacent meta-atoms is denoted by $r_{\text{atom}} $, while the length and width of each meta-atom are given by $d_{x}$ and $d_{y} $, respectively. The horizontal distance between the antenna array and the first metasurface layer is denoted by $d_{t}$, while the spacing between each metasurface layer is denoted by $d_{\text{Layer}}$. \par
Furthermore, let $e^{j\theta _{n}^{l} }$ be the transmission coefficient of the $n$-th meta-atom on the $l$-th metasurface layer, with $\theta _{n}^{l}\in \left [0,2\pi \right )$. The matrix of transmission coefficients for the $l$-th metasurface layer is given by $\mathbf{\Phi}^{l}=\text{diag}\left ( e^{j\theta _{1}^{l}},e^{j\theta _{2}^{l}},\dots ,e^{j\theta _{N}^{l}} \right )\in \mathbb{C}^{N\times N}$ \cite{di2020smart}. The matrix of propagation channels from the $\left ( l-1 \right ) $-th to the $l$-th metasurface layer is denoted by $\mathbf{W}^{l} \in \mathbb{C}^{N\times N}, \forall l\ne 1, l\in\mathcal{L}$. According to Rayleigh-Sommerfeld's diffraction theory \cite{All, arXiv_2024_An_Two}, the $(n,n')$-th entry of $\mathbf{W}^{l}$ can be written as
\begin{equation}
\setlength{\abovedisplayskip}{2pt}
\setlength{\belowdisplayskip}{2pt}
\label{1}
   w^{l}_{n,n'}= \frac{d_{x}d_{y}d_{\text{Layer} }}{{d^{l}_{n,n'}}^2}   \left(\frac{1}{2\pi d^{l}_{n,n'}}-\frac{1}{\lambda }j \right)e^{j2\pi d^{l}_{n,n'}/\lambda  },  
\end{equation}
where $\lambda$ is the wavelength, $d^{l}_{n,n'}$ is the transmission distance from the $n'$-th meta-atom of the $\left ( l-1 \right )$-th metasurface layer to the $n$-th meta-atom of the $l$-th metasurface layer. The vector of propagation channels from the $m$-th antenna to the first metasurface layer is denoted by $\mathbf{w}_{m} \in \mathbb{C}^{N\times 1}$ and can be obtained by replacing $d^{l}_{n,n'}$ in (\ref{1}) with $r_{n,m}^{1}$, as shown in Fig. \ref{Figure1b}.\par
The SIM-enabled wave-based beamforming matrix can be written as $\mathbf{G}=\mathbf{\Phi}  ^{L} \mathbf{W}^{L}\cdots  \mathbf{\Phi}  ^{2}\mathbf{W}^{2}    \mathbf{\Phi}  ^{1} \in \mathbb{C}^{N\times N}.$
Moreover, we assume a quasi-static flat fading channel model, i.e., the channel parameters within a coherent block are time-invariant \cite{Stacked1,tse2005fundamentals}. The baseband equivalent channel from the last metasurface layer to the $k$-th UE is denoted by $\mathbf{h}_{k}^{H} \in \mathbb{C}^{1\times N}, \forall k\in\mathcal{K}$, which is modeled by $\mathbf{h}_{k} \sim \mathcal{CN} \left(\mathbf{0},\beta _{k}\mathbf{R} \right)$, where $\beta _{k} $ denotes the distance-dependent path loss between the last metasurface layer and the $k$-th UE, $\mathbf{R}\in \mathbb{C}^{N\times N}$ denotes the covariance matrix characterizing the spatial correlation between each meta-atom on the last metasurface layer. According to \cite{Channel}, we can write the $\left (n,n' \right)$-th entry of $\mathbf{R}$ as
\begin{equation}
\setlength{\abovedisplayskip}{2pt}
\setlength{\belowdisplayskip}{2pt}
\begin{aligned}
\label{4}
    \left [ \mathbf{R} \right ]_{n,n'}=&\iint_{-\pi /2}^{\pi /2} f \left (\varphi ,\theta \right ) \\
    &\times e^{j2\pi \left (d_{n,n'}^{H}\sin \left (\varphi \right )\cos \left (\theta \right )+d_{n,n'}^{V}\sin \left (\theta \right) \right )/\lambda }d\theta d\varphi, 
\end{aligned}
\end{equation}
where $\varphi$, $\theta$, $f \left (\varphi ,\theta \right )$, $d_{n,n'}^{H}$ and $d_{n,n'}^{V}$ denote the azimuth angle, elevation angle, spatial scattering function, the horizontal and vertical distances between the $n$-th and $n'$-th meta-atoms on the same metasurface layer, respectively. We assume that all metasurface layers have a square structure, and the metasurface centers are in the same horizontal line. Furthermore, all layers are parallel to each other, and the transmit antenna array is parallel to the SIM.
\vspace{-0.25cm}
\section{Channel Estimation for SIM-Assisted Multi-User HMIMO Systems}
\vspace{-0.05cm}
\label{CHANNEL}
In this section, we investigate the channel estimation problem in SIM-assisted multi-user systems. As customary, we assume that the number of meta-atoms $N$ on each metasurface layer is generally larger than the number of antennas $M$ at the BS \cite{Stacked1}. Additionally, the SIM cannot probe the wireless channels associated with each meta-atom individually, making it difficult to estimate the channels from the UEs to the last layer of the SIM in practice. To overcome this challenge, we consider collecting multiple copies of the pilot signals to probe the uplink channel. \par
\begin{figure*}[!htp]
\centering 
\includegraphics[width=17cm]{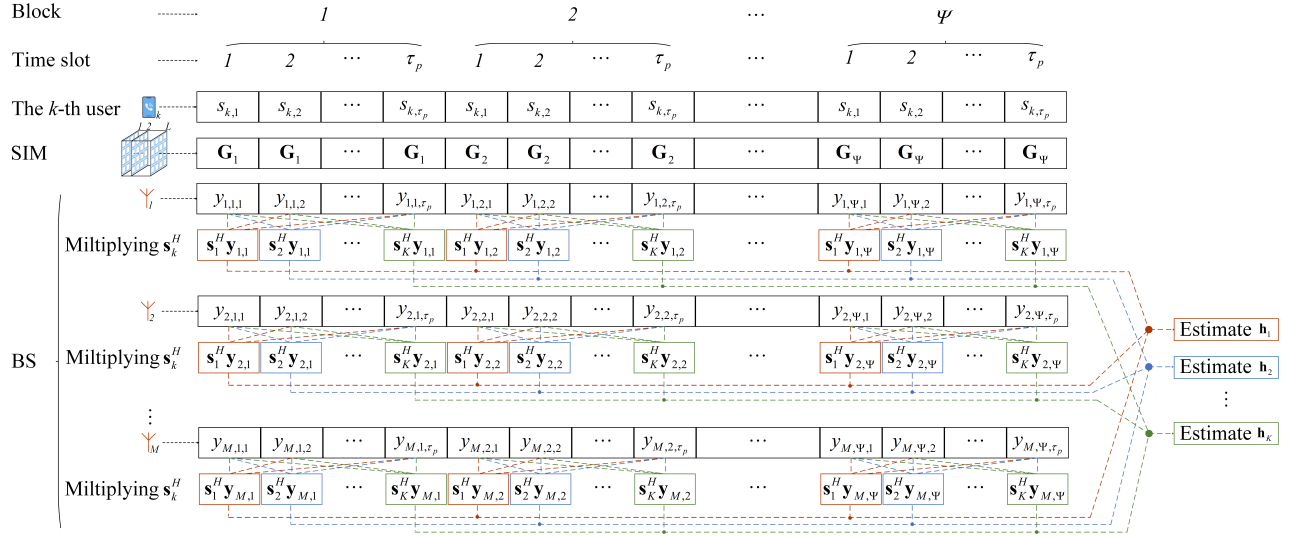} 
\caption{The proposed channel estimation protocol.}
\label{Figure3}
\vspace{-0.55cm}
\end{figure*}
As shown in Fig. \ref{Figure3}, we consider $\Psi$ blocks, each containing $\tau _{p}$ time slots. All the UEs send the same pilot sequence of length $\tau _{p}$ over different blocks. Let $s_{k,t}$ represent the pilot signal sent by the $k$-th UE in the $t$-th time slot of each block. The average power of the pilot signal of the $k$-th UE is denoted by $p_{k}$. Therefore, the signal received by the $m$-th antenna in the $t$-th time slot of the $\psi$-th block can be written as
\begin{equation}
\setlength{\abovedisplayskip}{2pt}
\setlength{\belowdisplayskip}{2pt}
\label{10}
    \begin{aligned}
         y_{m,\psi,t}=\mathbf{w}_{m}^{H}\mathbf{G}_{\psi}^{H}\sum_{k'=1}^{K}\mathbf{h}_{k'} \sqrt{p_{k'}}s_{k',t}+n_{m,\psi,t},
    \end{aligned}
\end{equation}
where $\mathbf{G}_{\psi}\in \mathbb{C}^{N\times N}$ denotes the beamforming matrix of the SIM in the $\psi$-th block, and $n_{m,\psi,t}\sim \mathcal{CN}\left (0,\sigma _{n}^2 \right )$ denotes the additive white Gaussian noise at the $m$-th antenna in the $t$-th time slot of the $\psi$-th block. For simplicity, we assume the same noise power $\sigma _{n}^2$ for all the antennas. As shown in Fig. \ref{Figure3}, by collecting the signals received at the $m$-th antenna in the $\psi$-th block, we obtain
\begin{equation}
\setlength{\abovedisplayskip}{2pt}
\setlength{\belowdisplayskip}{2pt}
\label{11}
\begin{aligned}
\underset{\mathbf{y}_{m,\psi}}{\underbrace{\begin{bmatrix}
y_{m,\psi,1}\\
y_{m,\psi,2}
 \\\vdots
 \\y_{m,\psi,\tau _{p}}
\end{bmatrix}}}\!=\! \!\sum_{k'=1}^{K} \! \sqrt{p_{k'}}\!\underset{\mathbf{s}_{k'}}{\underbrace{\begin{bmatrix}
s_{k',1}\\
s_{k',2}
 \\\vdots
 \\s_{k',\tau _{p}}
\end{bmatrix}}}\!\mathbf{w}_{m}^{H}\mathbf{G}_{\psi}^{H}\mathbf{h}_{k'}
\!+\!
\underset{\mathbf{n}_{m,\psi}}{\underbrace{\begin{bmatrix}
n_{m,\psi,1}\\
n_{m,\psi,2}
 \\\vdots
 \\n_{m,\psi,\tau _{p}}
\end{bmatrix}}},
\end{aligned}
\end{equation}
where $\mathbf{y}_{m,\psi} \in \mathbb{C}^{\tau _{p}\times 1}$ and $\mathbf{n}_{m,\psi}\in \mathbb{C}^{\tau _{p}\times 1}$ denote the received signal and noise vector by the $m$-th antenna, respectively, in the $\psi$-th block. The pilot sequence sent by the $k$-th UE is $\mathbf{s}_{k}\in \mathbb{C}^{\tau _{p}\times 1}$. To eliminate the interference between multiple UEs in each block, we adopt the mutually orthogonal pilot sequence strategy, such that $\sum_{t=1}^{\tau _{p}}s_{k,t}s_{k',t} =\tau_{p}$, for $k=k'$ and $ \sum_{t=1}^{\tau _{p}}s_{k,t}s_{k',t} =0$, for $k\ne k'$. One efficient pilot design is to employ the columns of the $\tau _{p}$-point discrete Fourier transform (DFT) matrix as the pilot sequences for different UEs. Note that the length of $\mathbf{s}_{k}$ needs to satisfy $\tau _{p}\ge K$ in order to separate the pilot signals sent by $K$ UEs within a block. In this letter, to minimize the pilot overhead, we consider $\tau_p = K$ in each block. Additionally, we note that the random phase drift due to practical hardware imperfections can be effectively solved by a compensation technique \cite{wei2022accurate}, and is ignored. After left-multiplying $\mathbf{s}_{k}^{H}$ on both sides of (\ref{11}) and collecting the output signals associated to the $M$ antennas, we arrive at
\begin{equation}
\label{13}
 \underset{\breve{\mathbf{y}}_{k,\psi}}{\underbrace{\begin{bmatrix}
\mathbf{s}_{k}^{H}\mathbf{y}_{1,\psi}
 \\\mathbf{s}_{k}^{H}\mathbf{y}_{2,\psi}
 \\\vdots 
 \\\mathbf{s}_{k}^{H}\mathbf{y}_{M,\psi}
\end{bmatrix}}}
=\tau_{p}\sqrt{p_{k}}
\underset{{\breve{\mathbf{W}}}}{\underbrace{\begin{bmatrix}
\mathbf{w}_{1}^{H}
 \\\mathbf{w}_{2}^{H}
 \\\vdots 
 \\\mathbf{w}_{M}^{H}
\end{bmatrix}}}
\mathbf{G}_{\psi}^{H}\mathbf{h}_{k}+
\underset{{\breve{\mathbf{n}}_{k,\psi}}}{\underbrace{\begin{bmatrix}
\mathbf{s}_{k}^{H}\mathbf{n}_{1,\psi}
 \\\mathbf{s}_{k}^{H}\mathbf{n}_{2,\psi}
 \\\vdots 
 \\\mathbf{s}_{k}^{H}\mathbf{n}_{M,\psi}
\end{bmatrix}}},
\end{equation}
where $\breve{\mathbf{y}}_{k,\psi}\in \mathbb{C}^{M\times 1}$ denotes the matched output signals, $\breve{\mathbf{W}}\in \mathbb{C}^{M\times N}$ denotes the channel propagation matrix from the first metasurface layer to the BS, and $\breve{\mathbf{n}}_{k,\psi}\in \mathbb{C}^{M\times 1}$ denotes the noise vector, in the $\psi$-th block. Upon collecting the matched output signals over $\Psi$ blocks, we obtain
\begin{equation}
\label{14}
\begin{aligned}
\underset{\tilde{\mathbf{y}}_{k}}{\underbrace{\begin{bmatrix}
\breve{\mathbf{y}}_{k,1}
\\\breve{\mathbf{y}}_{k,2}
 \\\vdots
 \\\breve{\mathbf{y}}_{k,\Psi}
\end{bmatrix}}} =\tau_{p}\sqrt{p_{k}}
\underset{\mathbf{P}}{\underbrace{\begin{bmatrix}
\breve{\mathbf{W}}\mathbf{G}_{1}^{H}
 \\\breve{\mathbf{W}}\mathbf{G}_{2}^{H}
 \\\vdots
 \\\breve{\mathbf{W}}\mathbf{G}_{\Psi}^{H}
\end{bmatrix}}}\mathbf{h}_{k}+
\underset{\tilde{\mathbf{n}}_{k}}{\underbrace{\begin{bmatrix}
\breve{\mathbf{n}}_{k,1}
 \\\breve{\mathbf{n}}_{k,2}
 \\\vdots
 \\\breve{\mathbf{n}}_{k,\Psi}
\end{bmatrix}}},
\end{aligned}          
\end{equation}
where $\tilde{\mathbf{y}}_{k} \in \mathbb{C}^{\Psi M \times 1}$, $\tilde{\mathbf{n}}_{k}\in\mathbb{C}^{\Psi  M\times1}$ and $ \mathbf{P}\in \mathbb{C}^{\Psi  M\times N}$ denote the received signal, the noise and the transmission matrix from the last metasurface layer to the BS, respectively, over $\Psi$ blocks. Furthermore, the variance of $\tilde{\mathbf{n}}_{k}$ is $\mathbb{E}\left \{ \left ( \tilde{\mathbf{n}}_{k} -\mathbb{E}\left \{ \tilde{\mathbf{n}}_{k} \right \} \right ) \left ( \tilde{\mathbf{n}}_{k} -\mathbb{E}\left \{ \tilde{\mathbf{n}}_{k} \right \} \right )^H   \right \}=\tau _{p}\sigma _{n}^2\mathbf{I}_{\Psi M}$, where $\mathbb{E}\left \{ \cdot  \right \}$ denotes the expectation, $\mathbf{I}_{\Psi M} \in \mathbb{C}^{\Psi M\times \Psi M}$ is the identity matrix. Based on the considered system model, the minimum number of blocks for the estimation of $\mathbf{h}_{k}$ is $\left \lceil N/M \right \rceil$. We consider $\Psi=\left \lceil N/M \right \rceil $ to minimize the pilot overhead. Based on (\ref{14}), we then focus on the estimation of $\mathbf{h}_{k}$.\par
Specifically, when $\mathbf{R}$ is known, the MMSE estimate of $\mathbf{h}_{k}$ is given by 
\begin{equation}
\setlength{\abovedisplayskip}{2pt}
	\setlength{\belowdisplayskip}{2pt}
{
    \label{MMSE}
   \hat{\mathbf{h}}_{k, \text{MMSE}}\!=\! \sqrt{p_{k}}\beta _{k}\mathbf{R}\mathbf{P}^H\!\left ( \! \tau_pp_{k}\beta _{k}\mathbf{P}\mathbf{R}\mathbf{P}^H \!+\! \sigma_{n}^2\mathbf{I}_{\Psi M} \!\right)\!^{-1}\tilde{\mathbf{y}}_{k},
}
\end{equation}
for $k\in\mathcal{K}$, where $\left \{ \cdot \right \}^{-1}$ denotes the inverse of a matrix. Although the MMSE estimator can suppress the noise, it requires the knowledge of the covariance matrix. If the matrix $\mathbf{R}$ is not known, on the other hand, the LS estimator can be utilized, yielding
\begin{equation}
\setlength{\abovedisplayskip}{2pt}
	\setlength{\belowdisplayskip}{2pt}
{
    \label{LS}
   \hat{\mathbf{h}}_{k, \text{LS}}=\mathbf{P}^{\dagger}\tilde{\mathbf{y}}_{k}/\left (  \tau_p\sqrt{p_{k}} \right ) , k\in \mathcal{K},
}
\end{equation}
where $\left \{ \cdot \right \}^{\dagger}$ denotes the pseudo-inverse of a matrix. Compared to the MMSE estimator, the LS estimator does not need any statistical information of $\mathbf{h}_k$, but it is very sensitive to the noise and its performance degrades in deep fading channels.\par
To address this issue, we then propose two subspace-based channel estimators, which relax the amount of information about $\mathbf{R}$ while achieving better performance than the LS estimator. By replacing $\mathbf{PR}\mathbf{P}^H$ with its singular value decomposition (SVD), i.e., $\mathbf{U}_{1}\mathbf{\Lambda} _{1}\mathbf{U}_{1}^H$, (\ref{MMSE}) can be rewritten as
\begin{equation}
{
\label{18}
    \begin{aligned}
\hat{\mathbf{h}}_{k, \text{MMSE}}&=\frac{\mathbf{P}^{-1}\mathbf{P}\mathbf{R}\mathbf{P}^H\left (\mathbf{P}\mathbf{R}\mathbf{P}^H+\frac{\sigma_{n}^2\mathbf{I}_{\eta }}{\tau_pp_{k}\beta _{k} } \right)^{-1}\tilde{\mathbf{y}}_{k}}{\tau_p\sqrt{p_{k}}} \\
&=\frac{\mathbf{P}^{-1}\mathbf{U}_{1}\left (\mathbf{\Lambda} _{1}\left (\mathbf{\Lambda} _{1}+\frac{\sigma_{n}^2\mathbf{I}_{\eta }}{\tau_pp_{k}\beta _{k} } \right )^{-1}\right )\mathbf{U}_{1}^H\tilde{\mathbf{y}}_{k}}{\tau_p\sqrt{p_{k}}},
    \end{aligned}
}
\end{equation}
where the diagonal matrix $\mathbf{\Lambda} _{1}\in \mathbb{C}^{\eta \times \eta}$, $\mathbf{U}_{1}\in \mathbb{C}^{\Psi M\times \eta}$ and $\eta$ represent the non-zero eigenvalues, the matrix constituted by the corresponding orthonormal eigenvectors, and the rank of $\mathbf{PRP}^H$, respectively. The invertibility of the transmission matrix $\mathbf{P}$ can be readily achieved by adjusting the phase shifts of the SIM. We notice that, as $p_{k}\to \infty $, $\left (\mathbf{\Lambda} _{1}\left (\mathbf{\Lambda} _{1}+\frac{\sigma_{n}^2\mathbf{I}_{\eta }}{\tau_pp_{k}\beta _{k} } \right )^{-1}\right )\to \mathbf{I}_{\eta }$. Leveraging this asymptotic approximation, the proposed RSLS estimator is given by
\begin{equation}
{
\label{19}
    \hat{\mathbf{h}}_{k,\text{RSLS}}=\mathbf{P}^{-1}\mathbf{U}_{1}\mathbf{U}_{1}^H\tilde{\mathbf{y}}_{k}\left /(  \tau_p\sqrt{p_{k}} \right ) ,k\in \mathcal{K},
}
\end{equation}
which is similar to the MMSE for large values of $p_{k}$. The RSLS estimator leverages the rank deficiency of the spatial correlation matrix $\mathbf{R}$ resulting from the reduced meta-atom spacing in HMIMO systems \cite{Channel}. Specifically, it projects the $\Psi M$-dimensional received signal $\tilde{\mathbf{y}}_{k}$ onto a low-dimensional space, effectively eliminating the noise caused by other $\left(\Psi M-\eta\right)$ dimensions. Hence, it achieves improved estimation accuracy compared to the LS estimator. Compared to the MMSE estimator, the RSLS estimator requires only prior information about the slowly varying matrix $\mathbf{U}_{1}$ which can be more readily obtained compared with $\mathbf{R}$.\par
However, $\mathbf{U}_{1}$ in (\ref{19}) still relies on prior knowledge of the considered scattering environment. When considering an isotropic scattering environment, the $(n,n')$-th entry of the spatial correlation matrix $\mathbf{R}_{\text{iso}}$ is $\left [ \mathbf{R}_{\text{iso}} \right ]_{n,n'} =\text{sinc}\left ( 2\sqrt{{d_{n,n'}^{H}}^{2}+{d_{n,n'}^{V}}^{2}}/\lambda   \right ) $ \cite{an2023tutorial,Rayleigh}. The span of the isotropic channel, i.e., $f\left (\varphi ,\theta \right )>0$, is across the entire angular domain, including the span of arbitrary practical channels \cite{Channel}. More specifically, the subspace generated by the columns of $\mathbf{R}_{\text{iso}}$ includes the subspace generated by the columns of more general matrices $\mathbf{R}$. Hence, we can utilize $f\left (\varphi ,\theta \right )$ for isotropic channels by employing the information about $\mathbf{R}_{\text{iso}}$. Taking this into account, the RSLS estimator can be rewritten as
\begin{equation}
{
\label{20}
    \hat{\mathbf{h}}_{k,\text{RSLS-iso}}=\mathbf{P}^{-1}\mathbf{U}_{2}\mathbf{U}_{2}^H\tilde{\mathbf{y}}_{k}/\left ( \tau_p\sqrt{p_{k}} \right )  , k\in \mathcal{K},
}
\end{equation}
where $\mathbf{U}_{2}\in \mathbb{C}^{\Psi M\times \eta_{\text{iso}}}$ is obtained by computing the SVD, i.e., $\mathbf{P}\mathbf{R}_{\text{iso}}\mathbf{P}^H=\mathbf{U}_{2}\mathbf{\Lambda} _{2}\mathbf{U}_{2}^H$, and $\eta_{\text{iso}}$ represents the rank of $\mathbf{P}\mathbf{R}_{\text{iso}}\mathbf{P}^H$. Accordingly, we can use the statistical channel state information (CSI) based on the array arrangement to achieve more accurate estimates of $\mathbf{h}_k$ than the LS estimator.
\vspace{-0.2cm}
\section{MSE of the Proposed Channel Estimators}
\label{COMPARISON}
In this section, we compute the mean square error (MSE) of the proposed LS, MMSE, RSLS, and RSLS-iso estimators. According to \cite{biguesh2006training}, the MSE of the MMSE and LS channel estimators are given by
\begin{align}
\setlength{\abovedisplayskip}{2pt}
	\setlength{\belowdisplayskip}{2pt}
\label{21}
\text{MSE}_{\text{MMSE}}&\!=\!\sigma_{n}^2\text{tr} \left \{ \beta _{k}\mathbf{R}\left (\tau_pp_{k}\beta _{k}\mathbf{P}^{H}\mathbf{PR}\!+\! \sigma_{n}^2\mathbf{I}_{N} \right )^{-1} \right \},\\
\label{22}
\text{MSE}_{\text{LS}}&=\sigma_{n}^2\text{tr} \left \{ \left (\tau_pp_{k}\mathbf{P}^H\mathbf{P} \right )^{-1} \right \} , 
\end{align}
where $\text{tr}\left \{ \cdot  \right \}$ denotes the trace. Furthermore, the MSE of the RSLS estimator is given by
\begin{equation}
\setlength{\abovedisplayskip}{2pt}
	\setlength{\belowdisplayskip}{2pt}
{
\label{MSE_3}
    \begin{aligned}
        &\text{MSE}_{\text{RSLS}}=\mathbb{E}\left \{ \left \| \mathbf{h}_k-\hat{\mathbf{h}}_{k, \text{RSLS}} \right \|^{2} \right \}  \\ 
&= \mathbb{E}\left \{ \left ( \mathbf{h}_k-\hat{\mathbf{h}}_{k, \text{RSLS}} \right )^H\left ( \mathbf{h}_k-\hat{\mathbf{h}}_{k, \text{RSLS}} \right )  \right \}          \\
&=\mathbb{E}\left \{ \mathbf{h}_{k}^{H}\left ( \mathbf{I}_{N}-\mathbf{X}-\mathbf{X}^{H}+\mathbf{X}^{H}\mathbf{X} \right )\mathbf{h}_{k}  \right \} \\
&+\mathbb{E}\left \{\left ( \mathbf{P}^{-1}\mathbf{U}_{1}\mathbf{U}_{1}^{H}\tilde{\mathbf{n}}_{k} \right )^{H} \mathbf{P}^{-1}\mathbf{U}_{1}\mathbf{U}_{1}^{H}\tilde{\mathbf{n}}_{k}\right\} /\left( \tau_p^{2}p_{k} \right), \\ 
\end{aligned}
}
\end{equation}
where $\mathbf{X}=\mathbf{P}^{-1}\mathbf{U}_{1}\mathbf{U}_{1}^{H}\mathbf{P}$. Specifically, the first entry in (\ref{MSE_3}) can be rewritten as
\begin{equation}
\setlength{\abovedisplayskip}{2pt}
	\setlength{\belowdisplayskip}{2pt}
\begin{aligned}
&\mathbb{E}\left \{ \mathbf{h}_{k}^{H}\left ( \mathbf{I}_{N}-\mathbf{X}-\mathbf{X}^{H}+\mathbf{X}^{H}\mathbf{X} \right )\mathbf{h}_{k}  \right \}\\
&=\beta _{k} \text{tr}\left \{\mathbf{R}\left ( \mathbf{I}_{N}-\mathbf{X}-\mathbf{X}^{H}+\mathbf{X}^{H}\mathbf{X} \right ) \right  \} \\
&=\beta _{k} \text{tr}\left \{ \left ( \mathbf{I}_{N}-\mathbf{X} \right )\mathbf{R}\left ( \mathbf{I}_{N}-\mathbf{X}^{H} \right )  \right \}\\
&=0. 
\end{aligned}
\end{equation}
Therefore, (\ref{MSE_3}) can be simplified as
\begin{equation}
\begin{aligned}
    \text{MSE}_{\text{RSLS}}=\frac{\sigma_{n}^2}{\tau_pp_{k}} \text{tr}\left \{ \mathbf{P}^{-1}\mathbf{U}_{1}\left ( \mathbf{P}^{-1}\mathbf{U}_{1} \right )^H  \right \}.
\end{aligned}    
\end{equation}
Similarly, the MSE of the RSLS-iso estimator is given by
\begin{equation}
\label{MSE_4}
    \begin{aligned}
        \text{MSE}_{\text{RSLS-iso}}=\frac{\sigma_{n}^2}{\tau_pp_{k}} \text{tr}\left \{ \mathbf{P}^{-1}\mathbf{U}_{2}\left ( \mathbf{P}^{-1}\mathbf{U}_{2} \right )^H  \right \}.
    \end{aligned}
\end{equation}
Moreover, we note that the MSE of these four estimators are highly dependent on the transmission matrix $\mathbf{P}$. Hence, the MSE of the considered estimators can be reduced by optimizing the phase shifts of the SIM during the training stage. In mathematical terms, the optimization of the SIM can be formulated as follows:
\begin{equation}
\setlength{\abovedisplayskip}{2pt}
	\setlength{\belowdisplayskip}{2pt}
\begin{aligned}
\label{target1}
 \underset{\theta _{n,\psi}^{l}}{\text{min}}\ &\text{MSE} \\
 s.t. \quad& \mathbf{P}=\left [ \mathbf{G}_{1}\breve{\mathbf{W}}^{H},\mathbf{G}_{2}\breve{\mathbf{W}}^{H},\cdots,\mathbf{G}_{\Psi}\breve{\mathbf{W}}^{H}  \right ]^{H},\\
 \ \quad& \mathbf{G}_{\psi }=\mathbf{\Phi}_{\psi }^{L} \mathbf{W}^{L}\cdots  \mathbf{\Phi}_{\psi }^{2}\mathbf{W}^{2}    \mathbf{\Phi}_{\psi }^{1},\\
 \ \quad& \mathbf{\Phi}_{\psi }^{l}=\text{diag}\left ( e^{j\theta _{1,\psi}^{l}},e^{j\theta _{2,\psi}^{l}},\cdots ,e^{j\theta _{N,\psi}^{l}}\right ), l \in \mathcal{L},\\
\ \quad& \theta _{n,\psi}^{l}\in \left [ 0,2\pi  \right ),l \in \mathcal{L},n \in \left \{ 1,2,\cdots ,N \right \}.
\end{aligned}
\end{equation}
Since the optimal configuration of the SIM involves a non-convex optimization problem \cite{nadeem2023hybrid}, in this paper, we consider a codebook-based scheme to heuristically produce a suboptimal solution of the phase shifts \cite{WC_2024_An_Codebook}. Specifically, for each estimator, we randomly generate multiple sets of phase shift vectors as the codebook and select the one that minimizes the corresponding MSE to configure the SIM during channel estimation.
\vspace{-0.2cm}
\section{Numerical Results}
\label{Simulation}
In this section, we illustrate simulation results to evaluate the performance of the proposed channel estimators. The actual channel coefficients are obtained as $\mathbf{h}_{k}=\sqrt{\beta_{k}} \mathbf{R}^{\frac{1}{2} }\mathbf{v}$, with $ \mathbf{v}\sim \mathcal{CN}\left (\mathbf{0},\mathbf{I}_{N} \right )$, and $ \beta _{k}=C_{0}d_{k}^{-\alpha }$, where $C_{0}$ is the path loss at a reference distance of $\SI{1}{m}$, $d_{k}$ is the distance from the $k$-th UE to the center of the last metasurface layer, and $\alpha$ is the path loss exponent \cite{Low}. We assume that $K$ UEs are randomly located within a disk parallel to the metasurface layer. The horizontal distance between the last metasurface layer and the disk is $\SI{25}{m}$, and the radius of the disk is $\SI{10}{m}$. Moreover, the carrier frequency is $\SI{28}{GHz}$. As for the large-scale path loss, we consider $ C_{0}=\SI{-30}{dB}$, $\alpha=2.8$, and the noise power is $\sigma_{n}^2=\SI{-104}{dBm}$. For the SIM, we consider $d_{t}=d_{\text{Layer}}=5\lambda/L$, $d_{x}=d_{y}=\lambda/8$ and $r_{\text{atom}}=\lambda/4$. The number of metasurface layers and the number of antennas are $L=6$ and $M=5$, respectively.\par
\begin{figure}[!t] 
\vspace{-0.3cm}
\centering
\subfloat[\tiny NMSE of estimators versus SNR.]  
{\label{Figure5}\includegraphics[width=7cm]{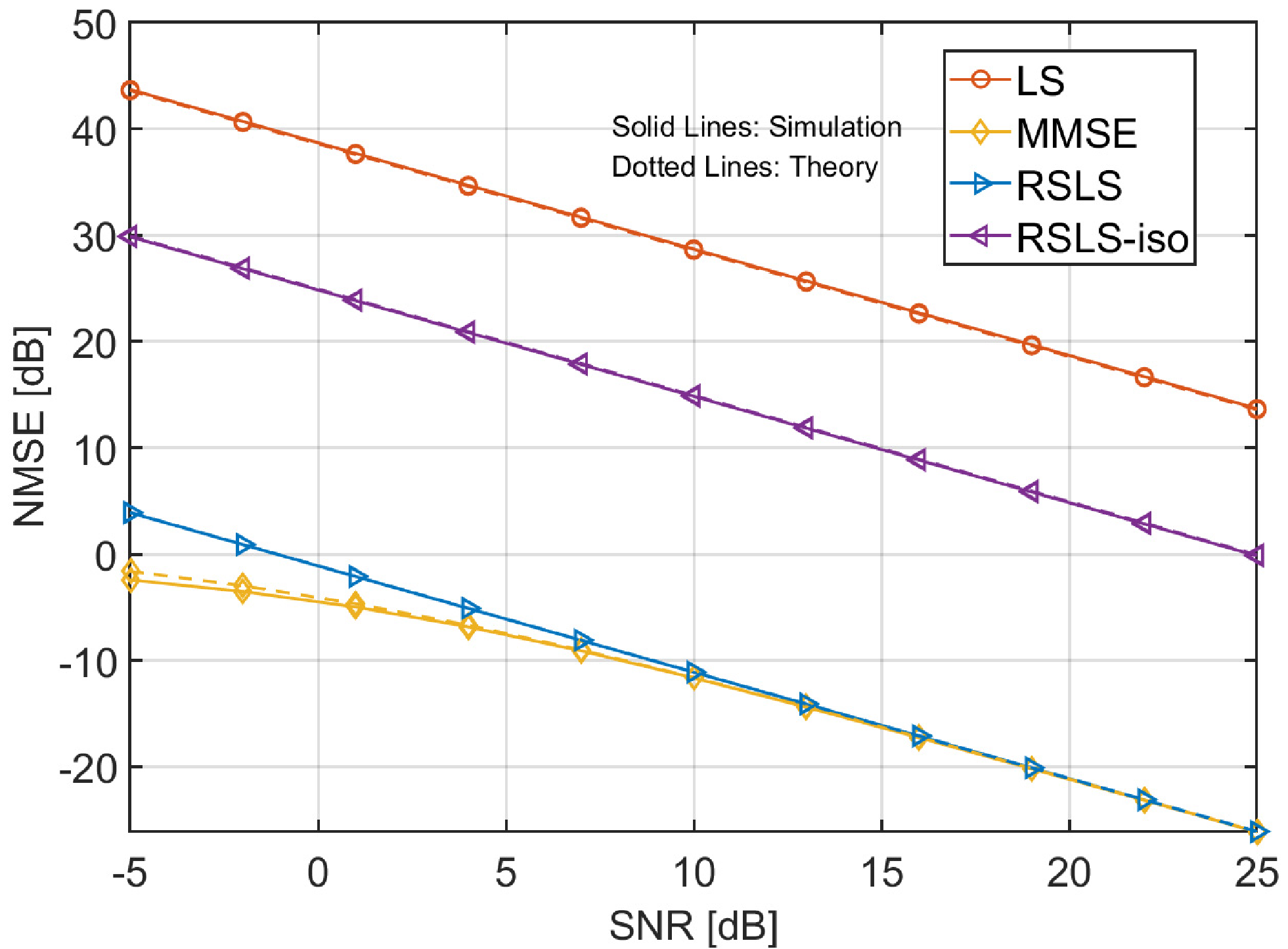}}
\subfloat[\tiny NMSE of estimators with $K=5$, $L=3$; $k=5$, $L=6$ and $K=20$, $L=6$.]
{\label{Figure6}\includegraphics[width=7cm]{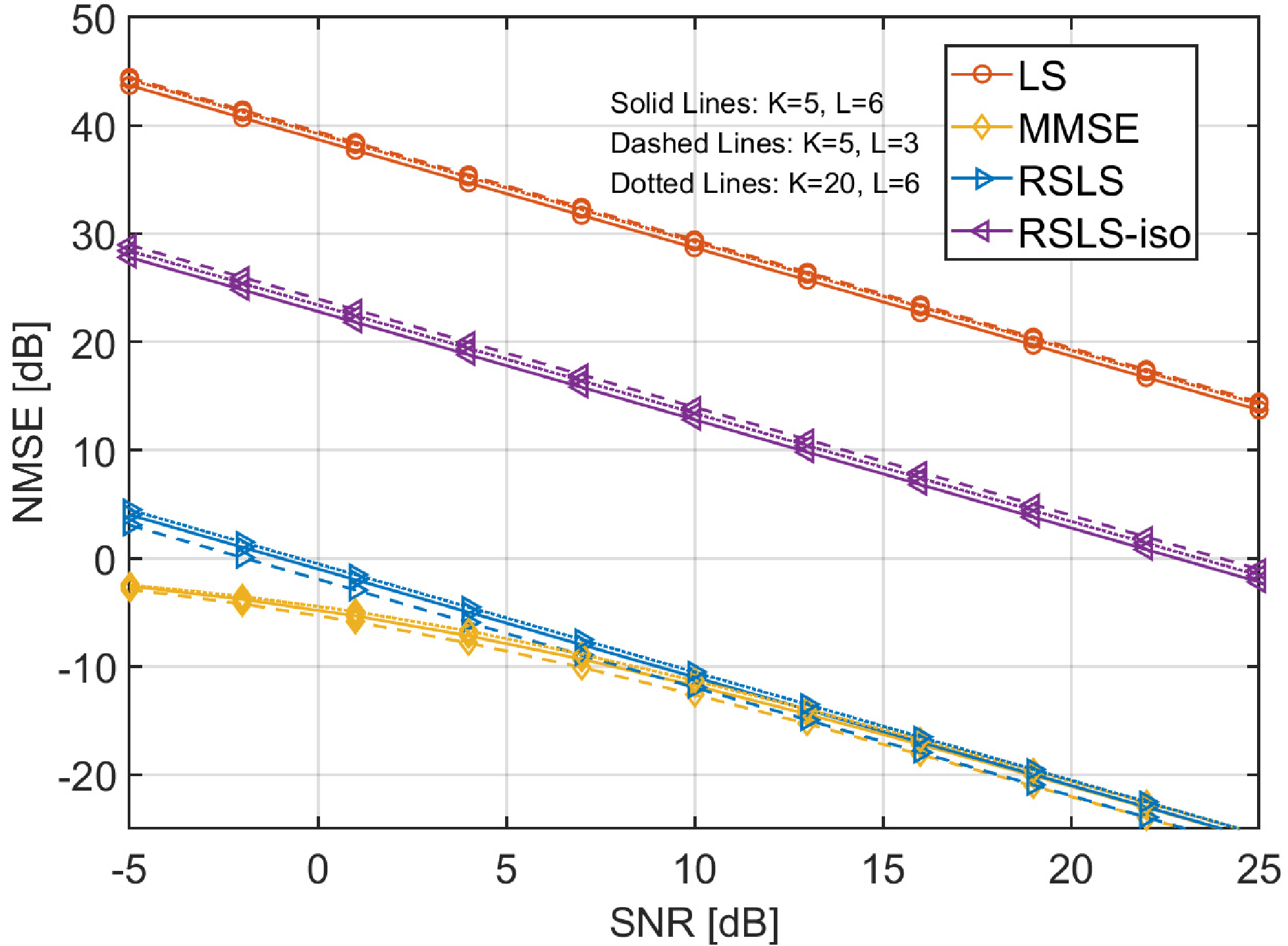}}
\vspace{-0.3cm}
\\       
\subfloat[\tiny NMSE of estimators with $M=10$, $N=100$; $M=5$, $N=100$ and $M=5$, $N=225$.]  
{\label{Figure7}\includegraphics[width=7cm]{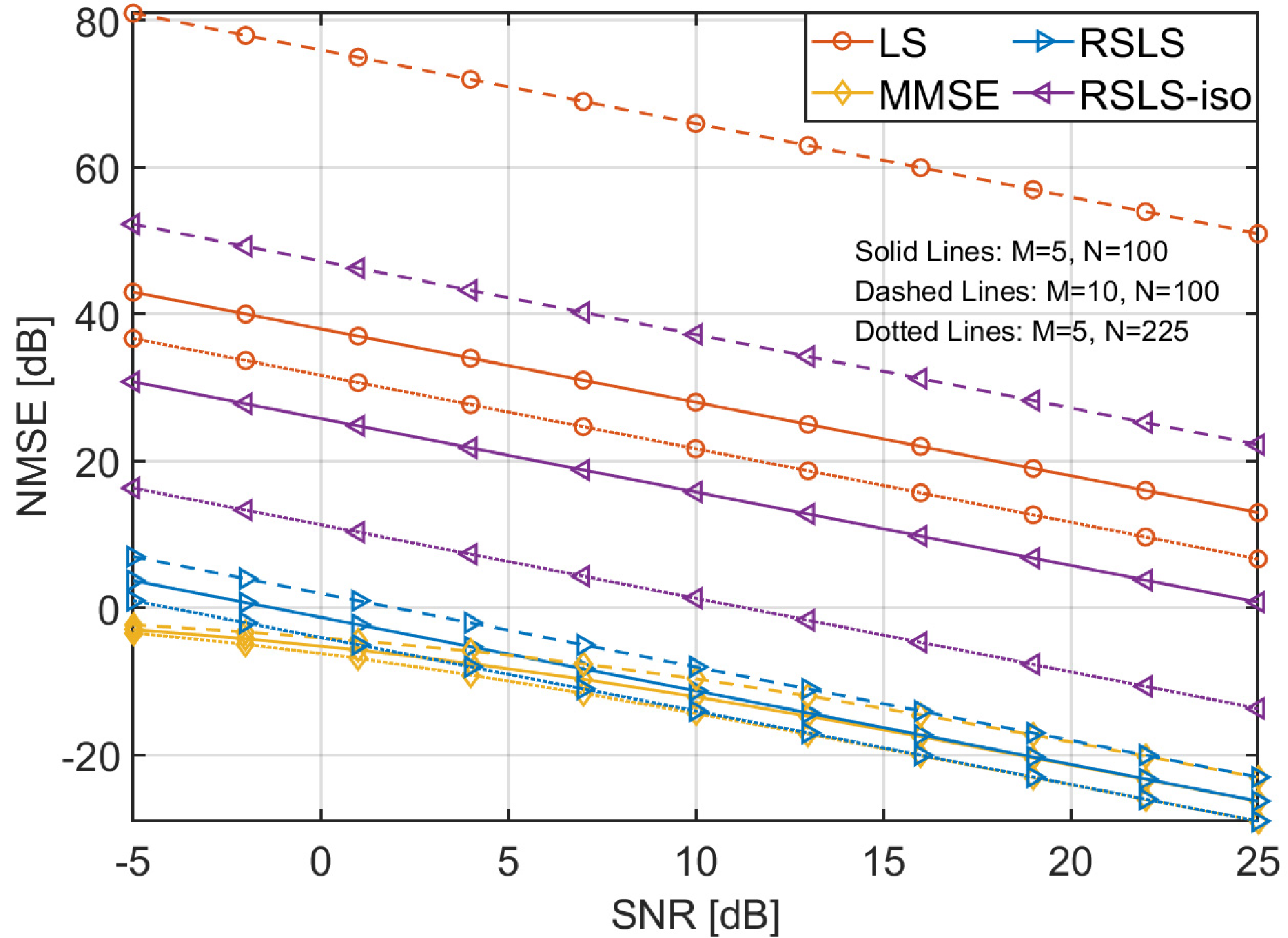}}
\subfloat[\tiny NMSE of estimators with random and optimal $\mathbf{P}$.]
{\label{Figure8}\includegraphics[width=7cm]{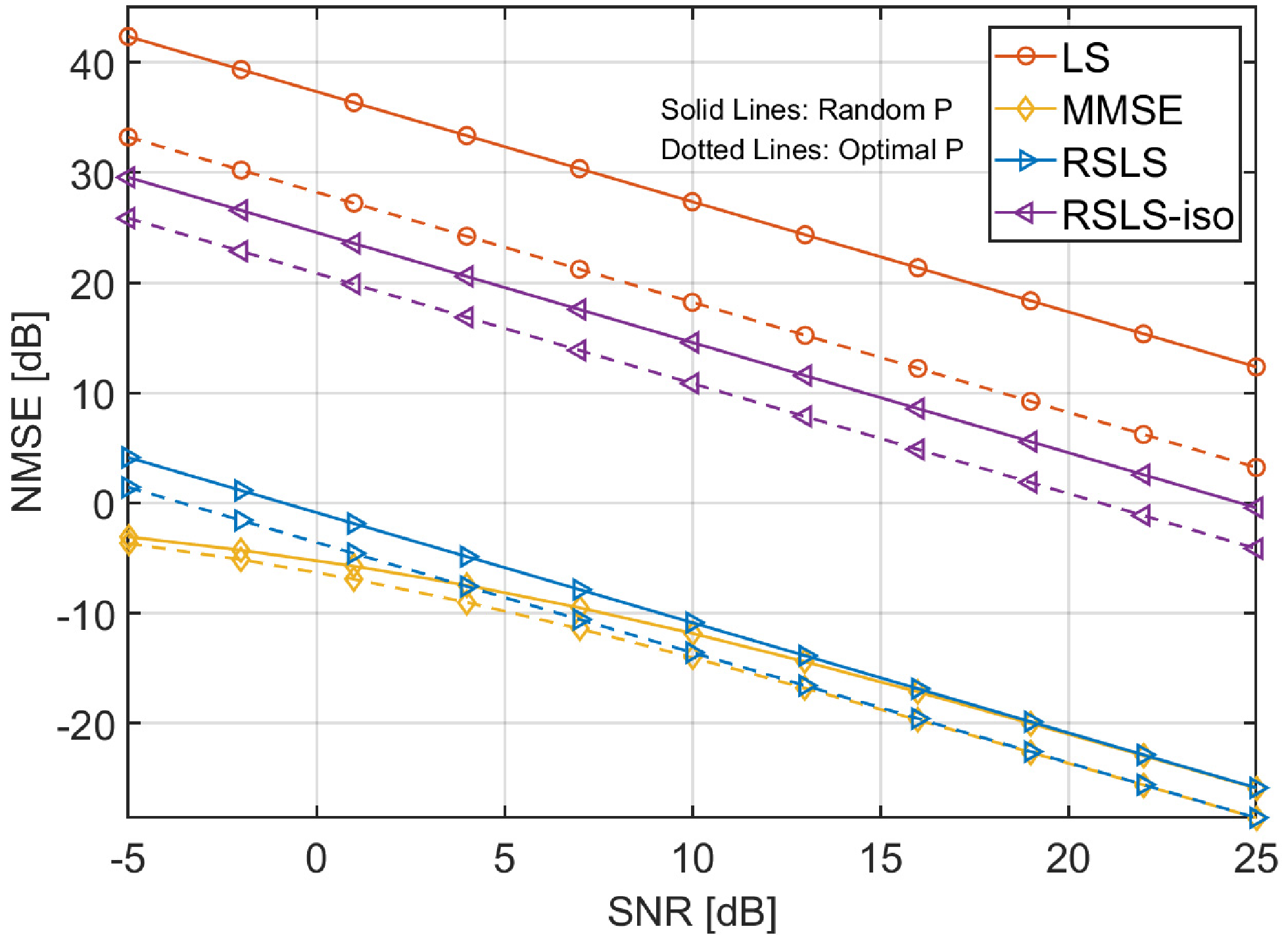}} 
\caption{NMSE of the proposed channel estimators.}    
\label{fig_4}    
\vspace{-0.7cm}
\end{figure}
First, we verify the analytical results by assuming that the spatial correlation matrix $\mathbf{R}_{\text{iso}}$ is known. As shown in Fig. \ref{Figure5}, the derived theoretical NMSEs accurately characterize the performance of all the proposed estimators. Furthermore, the LS estimator is highly sensitive to the noise, while the MMSE estimator can effectively suppress the noise and provide more than $\SI{35}{dB}$ performance gain. The RSLS estimator relying on statistical CSI offers the same performance as the MMSE estimator in the high SNR region. Moreover, although the RSLS-iso estimator underperforms the MMSE and RSLS estimators, it still provides a performance gain of $\SI{14}{dB}$ over the LS estimator by exploiting the rank deficiency of $\mathbf{R}_{\text{iso}}$.\par 
 In Fig. \ref{Figure6}, we evaluate the NMSE of the proposed channel estimators considering a different number of UEs and SIM layers. Note that the proposed channel estimation protocol can mitigate the interference among multiple UEs and is slightly affected by $L$. Moreover, Fig. \ref{Figure7} evaluates the impact of the number of antennas and meta-atoms on the estimation performance, which demonstrates that the NMSE of the proposed estimators is improved by increasing the number of meta-atoms $N$. This is because the rank deficiency of the spatial correlation matrix is more significant as the number of meta-atoms increases \cite{Channel}. Also, we increase the number of antennas $M$ from $5$ to $10$, and observe that the performance of all channel estimators degrades due to the fact that the properties of the matrix $\mathbf{P}$ are limited by the reduced number of observation blocks $\Psi=\left \lceil N/M \right \rceil $. Under all setups, the MMSE estimator and the RSLS estimator offer a stronger robustness than the LS estimator and the RSLS-iso estimator. Finally, we select the optimal $\mathbf{P}$ for each estimator from a random codebook of size $100$ that minimizes the MSE of the four estimators. Compared to a random SIM configuration, Fig. \ref{Figure8} indicates that a well-optimized SIM is capable of improving the NMSE for all the estimators.
\section{Conclusions}
\label{Conclusions}
In this letter, we proposed channel estimators for SIM-assisted multi-user HMIMO communication systems. Channel estimation is, in fact, a crucial task for subsequent beamforming and signal processing in SIM-assisted communication systems. To this end, we designed an effective channel estimation protocol to solve the underdetermined problem where the number of antennas at the BS is smaller than the number of meta-atoms on the last metasurface layer of the SIM. Four efficient channel estimation techniques were proposed. Additionally, we derived the MSE of the proposed channel estimators and minimized them by appropriately adjusting the phase shifts of the SIM. 
\ifCLASSOPTIONcaptionsoff
  \newpage
\fi
\bibliographystyle{IEEEtran}
\bibliography{IEEEabrv,Ref}
\end{document}